# Tangent functional connectomes uncover more unique phenotypic traits

Kausar Abbas[1,2,*], Mintao Liu[1,2,*], Michael Wang[1,2], Duy Duong-Tran[3,4], Uttara Tipnis[5], Enrico Amico[6,7], Alan D. Kaplan[5], Mario Dzemidzic[8], David Kareken[8], Beau M. Ances[9], Jaroslaw Harezlak[10], Joaquín Goñi[1,2,11,+].

[1]Purdue Institute for Integrative Neuroscience, Purdue University, West Lafayette, Indiana, USA.
[2]School of Industrial Engineering, Purdue University, West Lafayette, Indiana, USA.
[3]Perelman School of Medicine, University of Pennsylvania, Pennsylvania, USA.
[4]Department of Mathematics, United States Naval Academy, Annapolis, Maryland, USA.
[5]Lawrence Livermore National Laboratory, Livermore, California, USA.
[6]Institute of Bioengineering, Center for Neuroprosthetics, Ecole Polytechnique Federale de Lausanne, Lausanne, Switzerland.
[7]Department of Radiology and Medical Informatics, University of Geneva (UNIGE), Geneva, Switzerland.
[8]Department of Neurology, Indiana University School of Medicine, Indiana Alcohol Research Center, USA
[9]Department of Neurology, Washington University in Saint Louis, School of Medicine, St Louis, MO, USA
[10]Department of Epidemiology and Biostatistics, Indiana University, Bloomington, Indiana, USA.
[11]Weldon School of Biomedical Engineering, Purdue University, West Lafayette, Indiana, USA.
*These authors contributed equally to this work.
+ Corresponding author: jgonicor@purdue.edu

## Abstract

Functional connectomes (FCs) contain pairwise estimations of functional couplings based on pairs of brain regions activity derived from fMRI BOLD signals. FCs are commonly represented as correlation matrices that are symmetric positive definite (SPD) matrices lying on or inside the SPD manifold. Since the geometry on the SPD manifold is non-Euclidean, the inter-related entries of FCs undermine the use of Euclidean-based distances and its stability when using them as features in machine learning algorithms. By projecting FCs into a tangent space, we can obtain tangent functional connectomes (tangent-FCs), whose entries would not be inter-related, and thus, allow the use of Euclidean-based methods. Tangent-FCs have shown a higher predictive power of behavior and cognition, but no studies have evaluated the effect of such projections with respect to fingerprinting.

In this work, we hypothesize that tangent-FCs have a higher fingerprint than "regular" (i.e., no tangent-projected) FCs. Fingerprinting was measured by identification rates (ID rates) using the standard test-retest approach as well as incorporating monozygotic and dizygotic twins. We assessed: (i) *Choice of the Reference matrix $C_{ref}$*. Tangent projections require a reference point on the SPD manifold, so we explored the effect of choosing different reference matrices. (ii) *Main-diagonal Regularization*. We explored the effect of weighted main diagonal regularization[1]. (iii) *Different fMRI conditions*. We included resting state and seven fMRI tasks, (iv) *Parcellation granularities from 100 to 900 cortical brain regions (plus subcortical)*, (v) *Different distance metrics*. Correlation and Euclidean distances were used to compare regular FCs as well as tangent-FCs. (vi) fMRI *scan length* on resting state and when comparing task-based versus (matching scan length) resting-state fingerprint.

Our results showed that identification rates are systematically higher when using tangent-FCs. Specifically, we found: (i) Riemann and log-Euclidean matrix references systematically led to higher ID rates for all configurations assessed. (ii) In tangent-FCs, *Main-diagonal regularization prior to tangent space projection* was critical for ID rate when using Euclidean distance, whereas barely affected ID rates when using correlation distance. (iii) ID rates were dependent on condition and fMRI scan length. (iv) *Parcellation granularity* was key for ID rates in FCs, as well as in tangent-FCs with fixed regularization, whereas optimal regularization of tangent-FCs mostly removed this effect. (v) Correlation distance in tangent-FCs outperformed any other configuration of distance on FCs or on tangent-FCs across the "fingerprint gradient" (here sampled by assessing test-retest, Monozygotic twins and Dizygotic twins). (vi)

ID rates tended to be higher in task scans compared to resting-state scans when accounting for fMRI *scan length*.

In summary, we posit that FCs, when projected to a tangent space, display more unique phenotypic traits, and thus have greater potential for developing clinical biomarkers based on brain functional connectivity.

## Introduction

Using fMRI data, functional connectivity between two brain regions is usually estimated as the Pearson's correlation coefficient between their BOLD time-series. Subsequently, whole-brain functional connectivity patterns can be summarized in the form of a symmetric correlation matrix referred to as the Functional Connectome (FC)[2,3]. One of the most crucial steps in brain connectomics is the comparison of FCs across participants[1,4–9], fMRI conditions[10,11], mental states, or disease progression[12–14]. Canonically, such comparisons first vectorize the FC matrices, and then compute Pearson's correlation-based dissimilarity (or similarity)[6,8], or less frequently, Euclidean distance[15] between the vectorized FCs. From an algebraic standpoint, this implicitly assumes that FCs are Euclidean objects that lie in a *linear* high-dimensional space, where each element of the FC represents a dimension. However, in fact, FCs lie on or inside a high-dimensional *non-linear curved manifold* (or surface), called the *Symmetric Positive Definite (SPD) manifold* (Figure 1)[1,8].

Until recently, this important fact that FCs lie on or inside a non-linear manifold, has been scarcely considered by the neuroscientific community[1,8,16–22]. As a result, most analyses and frameworks did not take a full advantage of functional connectivity data to uncover their fingerprinting and/or biomarker capacity. This may have also limited the capacity of FC to predict cognitive outcomes or serve as reliable and robust clinical biomarkers of brain disorders. The use of Riemannian geometry may mitigate such limitations and enable comparing FCs with basic algebraic operations on the manifold when the underlying non-linear geometry of the correlation-based FCs is incorporated. In comparison to "regular" FCs[16,23], tangent-FCs have been proven to provide more accurate predictions of disease[19,21,24,25] and aging[26]. Also, Riemannian geometry-based approaches applied to functional connectivity have been recently used for harmonization of multi-site data[27], as well as for brain connectivity interface[28,29]. These findings pose a question about why tangent-FCs seem to be better biomarkers. We hypothesize that unsupervised transformations of FC data previously shown to improve associations related to cognition, behavior, and other biomarkers will have better fingerprinting. Our study fills that gap among tangent FCs, biomarkers, and the role of fingerprinting.

To some extent, FCs possess a recurrent and reproducible individual fingerprint[30–32] that can identify an individual from a population of FCs[1,6,7,33]. This process is referred to as "fingerprinting" or subject-identification. Using Pearson correlation coefficient as a similarity measure, individual fingerprints have been shown to exist in many different fMRI conditions (including resting state). Interestingly, aside from the resting-state fMRI data, identification rates were moderate to low in all other conditions[6]. Recently, Venkatesh et al.[8] argued that this is due to not accounting for the underlying geometry of FCs when they are compared. These authors proposed that the distance between FCs is better measured along the geodesic distance of the SPD manifold. They showed that by using geodesic distance instead of Pearson's correlation-based dissimilarity metrics, identification rates[6] increased robustly across a range of fMRI conditions. We recently extended this approach by showing the existence of an *optimal* amount of main diagonal regularization of FCs to maximize fingerprinting[1]. When such optimal regularization is applied to the FCs, it yields even higher identification rates for all available fMRI conditions in the Young-Adult HCP Dataset[34].

One limitation of the *Geodesic* and *optimally regularized Geodesic distance* is that it only provides a single numeric distance estimate between FCs, hence precluding element-wise (or edge-wise) analyses of FCs (i.e., focusing on a particular brain region or a specific functional coupling between two brain regions). Moreover, since FCs lie on or inside the SPD manifold, their individual elements (i.e., functional edges) are bound by the SPD criterion; that is, their entries are inter-related measurements[20]. As mentioned above, most functional connectivity frameworks do not incorporate this property and, instead, implicitly treat functional edges as independent or unrelated features. Many classifier algorithms are unstable in the face of inter-related features, i.e., small perturbations of the training data can alter the relative weighting of the features[35]. This may limit the generalizability of the classifiers and affect performance of machine learning algorithms. Furthermore, it makes the identification of significantly discriminative brain connections (from the classifier weights) non-trivial[36].

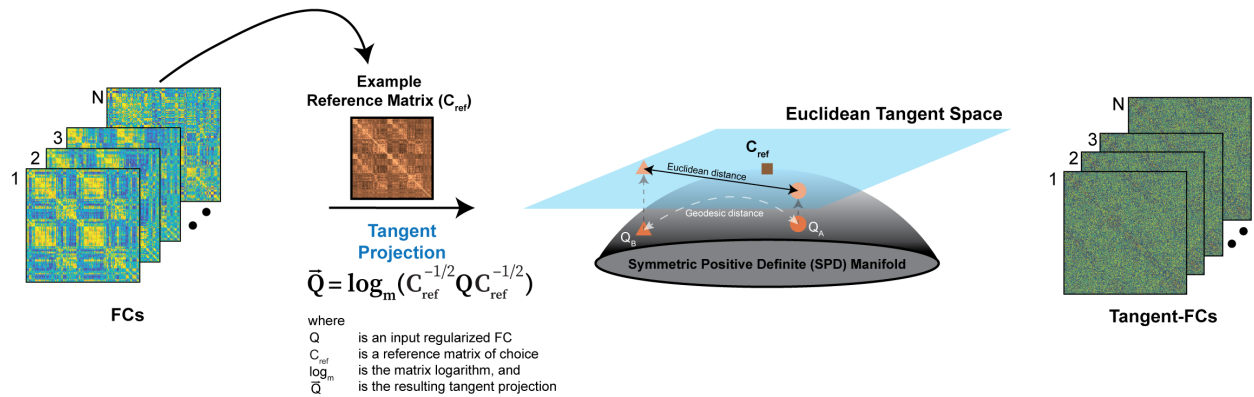

**Figure 1: Illustration of a Tangent Space Projection of FCs.** A sample of FCs is used to compute a reference matrix, $C_{ref}$, which is used as the point of the SPD manifold upon which the tangent space is created. Using $C_{ref}$ and the analytical formula for the tangent space projection, all FCs (being correlation matrices) can then be projected to a tangent-space.

This limitation can be addressed by using a Riemannian geometry tool. Briefly, this tool transforms a correlation matrix (which does not conform to Euclidean distances) and creates a new matrix (the tangent space projection of the original FC matrix) whose entries can be analyzed with Euclidean methods. That is, the projection of FCs from the SPD manifold onto a tangent space (Figure 1), which is Euclidean and permits the applications of Euclidean geometry methods[37,38]. It is noteworhty that a tangent space projection of a correlation matrix representing an FC, produces as outcome a tangent-FC. These tangent-projected values form a tangent-FC matrix of the same size as the input FC but with dimensionless connectivity units (see Figure 1 Right for examples on tangent-FCs) that constitute the new dependent variables to be analyzed. In a tangent-FC, elements (i.e., functional edges) are no longer bounded by the SPD criterion and edges can be treated individually, as they become unrelated features[20].

To date, no study has assessed the impact of FC fingerprinting when tangent space projections leading to unbounded edges are used. In this study, we explore the effects of tangent space projections of FCs from a fingerprinting standpoint. To do so, we assess how identification rates are affected by tangent-space projections of FCs for (i) fMRI conditions, (ii) parcellation granularities, (iii) magnitudes of main-diagonal regularizations, (iv) reference matrices used to project FCs, (v) distance used in a tangent space projection, and (vi) fMRI scan length. Furthemore, we build on the concept of fingerprinting gradient by including monozygotic and dizygotic twins under the conditions listed above. Finally, we tested the optimal settings found on an additional validation set (resting state).

## Materials and Methods

### HCP young-adult dataset

In this work, we used data from the HCP 1,200 participants release[34] and extracted three different subsets. The first consists of 426 unrelated participants (223 women, mean age: 28.67 years old, range: 22-36) selected so that no two participants have a shared parent, while the remaining subsets include 63 pairs of Monozygotic (MZ) twins and 63 pairs of Dizygotic (DZ) twins. To minimize shared environment differences in the MZ and DZ cohorts[39], only DZ twins of the same sex were included. This HCP dataset has more than 63 MZ pairs but we matched it to the DZ cohort size to make fingerprinting results easier to compare. fMRI data included both resting state (RS) and tasks: emotion processing (EM), gambling (GAM), language (LAN), motor (MOT), relational processing (REL), social cognition (SOC), and working memory (WM). For simplicity, we refer to resting state and the tasks as *fMRI conditions,* or simply *conditions*.

For each condition, participants underwent two sessions corresponding to two different acquisitions (left-to-right or LR, and right-to-left or RL). The resting-state scans were acquired on two different days for a total of four sessions ("REST1" and "REST2"). Only the two sessions from REST1 were utilized in this work. The HCP scanning protocol was approved by the institutional review board at Washington University in St. Louis. Full details on the HCP dataset have been published previously.

### Validation dataset

A cohort consisting of 181 participants was used to validate the effect of tangent space projection on ID rates (86 women, mean age: 34.59 years old, range: 19-89). The participants were healthy controls who enrolled in other than HCP studies between 2006 and 2020 at Washington University. For each participant, a structural T1-weighted scan and two runs of 6 mins resting-state fMRI scans were acquired using a 3T Siemens Tim-Trio MR scanner with a 12-channel head coil array. The T1-weighted scan was a high-resolution, 3D, sagittal, magnetization-prepared rapid gradient echo scan (MP-RAGE; repetition time (TR)=2400 ms; echo time (TE)=3.16 ms; flip angle=8 degrees; voxel size=1×1×1 mm$^3$; 256×256×176 acquisition matrix). Two resting-state fMRI runs were collected back-to-back using an echo planar sequence (voxel size=4 mm$^3$; repetition time=2200ms; Flip Angle=90 degrees).

### Brain parcellations

In this work, we have used a collection of functional brain atlases of the cortex, known as the *Schaefer parcellation*[40]. The Schaefer parcellation is based on resting-state fMRI data from 1,489 participants which were registered using surface alignment. To derive the Schaefer parcellation, a gradient-weighted Markov random field was employed which integrates local gradient and global similarity approaches. The Schaefer parcellation is available at ten granularity levels: 100-1000 in steps of 100. The Schaefer parcellations are available both in volumetric and grayordinate space. Since the grayordinate versions of the parcellations are in the same surface space as the HCP fMRI data, it is rather straightforward to map the parcellations onto the fMRI data. Furthermore, the alignment between the fMRI data and the Schaefer parcellations is much better when surface-mapping is used, as compared to the volumetric mapping. Hence, we used the surface-based mapping to map the 100-900 granularity Schaefer parcellations onto the fMRI data. At the time of the data processing for this study, we could not map the 1,000 Schaefer parcellation successfully for the HCP Young Adult dataset. For completeness, 14 subcortical regions were added to each parcellation, as provided by the HCP release (filename Atlas_ROI2.nii.gz). To do so, this file was converted from NIFTI to CIFTI format using the HCP workbench software

([www.humanconnectome.org/software/connectome-workbench.html](www.humanconnectome.org/software/connectome-workbench.html), *wb_command -cifti-create-label*)[41,42]. This resulted in, for example, a total 114 brain regions for the Schaefer-100 parcellation.

**Preprocessing of HCP dataset**

A "minimal" preprocessing pipeline from the HCP was employed[43], comprising artifact removal, motion correction, and registration to standard template. Full details can be found in earlier publications[43,44].

We added the following steps to the "minimal" pipeline. For resting-state fMRI data we: (i) regressed out the global gray matter signal from the voxel time courses[45], (ii) applied a first-order Butterworth bandpass filter in the forward and the reverse directions [$0.001-0.08 Hz$[45]; MATLAB functions *butter* and *filtfilt*], and (iii) z-scored and averaged, per brain regions, the voxel time courses, excluding any outlier time points falling outside three standard deviation from the mean (workbench software, *wb_command -cifti-parcellate*). For task-fMRI, we performed the same steps, but applied a more liberal frequency range was adopted for the bandpass filter (0.001-0.250)[46], since the relationship between different tasks and optimal frequency ranges is still unclear [47].

**Preprocessing of validation dataset**

This validation dataset was processed using an in-house pipeline based on AFNI[51,52], FSL[53], and Matlab using state-of-the-art guidelines. The same cortical parcellation scheme (Schaefer parcellation[40]) was used as we introduced in Methods section *Brain parcellations*, while subcortical regions were from scale I Tian parcellation[54].

Structural T1 images were first denoised to improve the signal-to-noise ratio (ANTs toolbox[55] *DenoiseImage*), bias-field corrected (*fsl_anat*), and then processed with the FreeSurfer (version 6) cortical reconstruction process (*recon-all*) to extract white matter, grey matter, and cerebrospinal fluid (CSF) tissue masks. For the cortical surface, the surface of the Schaefer parcellations were mapped from the template space to the native T1 surface space (*mri_surf2surf*) and then mapped onto a volume (*mri_aparc2aseg*). The mask for subcortical regions was obtained by running FIRST toolbox[56] to where Tian parcellations[54] were mapped using AFNI command (*auto_warp.py*).

To process the resting-state fMRI data, we modified the standard preprocessing pipeline from the "afni_proc.py" AFNI script. These steps included: removal of the first 4 TRs (*3dTcat*), computing outlier fraction for each volume (*3dToutcount*), removing spikes (*3dDespike*), performing slice timing correction (*3dTshift*), the registration of each volume to the base volume (*3dvolreg*), computing anatomical alignment transformation to EPI registration base (*align_epi_anat.py*), applying DiCER[57] (a novel method used to estimate a regressor as a correct measure of the GSR), blurring (*3dmerge*, 4.0mm Full Width at Half Maximum), and scaling (*3dcalc*). Regressors included six estimated head motion parameters and their derivatives, bandpass filtering (0.01 to 0.1 Hz, *1dBport*), tissue signals (3 principal components from ventricles and 5 from white matter), and DiCER[57]. All these nuisance regressors, along with censoring of volumes, were used to denoise the rs-fMRI data (*3dDeconvolve* and *3dTproject*).

**Estimation of whole-brain functional connectomes**

Pearson's correlation coefficient (MATLAB command *corr*) was used to estimate the functional connectivity between all pairs of brain regions, resulting in a symmetric correlation matrix of size $m \times m$, where $m$ is the number of brain regions for a given parcellation. We refer to this object as an FC. For each participant, we computed a whole-brain FC for each of the two sessions (also referred to as test-retest), each fMRI condition (resting state and all tasks), and each parcellation granularity (100-900).

**Quantification of fingerprinting**

For two paired samples of FCs (test-retest) of unrelated participants, *Fingerprinting* is the process of identifying an individual's FC from one session, given the FC of that individual from a second session. All conditions (resting state and seven tasks) in our dataset contain two runs (LR and RL acquisition). To minimize any acquisition orientation-based bias, each participant's sessions were randomly assigned to either test or retest. This process was repeated for each condition separately.

For each FC from the test sample, the most similar FC from the retest sample is identified. We then predict that the FC from the test sample must belong to the same participant than the most similar one from the retest sample (equivalent to a nearest neighbor classifier based on a specific similarity measure). The relative frequency of successful identifications is called "identification rate", and is obtained as:

$$\text{Identification (ID) Rate} = \frac{\text{Number of correctly labelled participants}}{\text{Total number of participants}}$$

Note that, in addition, this process must be done reversing the roles of test and retest sessions, as introduced by Finn and colleagues[6]. The final identification rate was obtained by averaging the two identification rate values (from test to retest and vice versa). We will also refer to this measure as "individual-fingerprinting".

For a sample of FCs of twin's data (MZ or DZ), the procedure was slightly different. Instead of selecting out of two sessions from the same individual, we take one session from one twin (twin1) and one session from the other twin (twin2). For each condition, each twin from a given pair has two runs (LR and RL). Once again to minimize biases due to the acquisition orientation, for each twin from a pair, a session was randomly selected and assigned to either twin1 or twin2 session. This process was repeated for each condition separately.

An FC from the twin2 session was labeled with the corresponding twin's identity in the twin1 data that was the closest to it in the twin1 session. We repeated this process for all the FCs in the twin2 session and ID rate in this case is defined as:

$$\text{Identification (ID) Rate} = \frac{\text{Number of correctly labelled twins}}{\text{Total number of twin pairs}}$$

Analogously to the test-retest protocol, this process was repeated by reversing the roles of the twin1 and twin2 sessions, and the final identification rate was obtained by averaging the two values. Note that we will also refer to it as "twin-fingerprinting", or "MZ/DZ-fingerprinting".

To assess variability of ID rates due to differences in samples, we used sampling without replacement. For every run, we randomly selected 80% of the participants and performed the fingerprinting process. This procedure was repeated 100 times and then the mean ID rate and standard error of the mean ID rate was computed. This "sampling without replacement" process also served as a proxy exploration of the generalizability of results obtained for other datasets acquired with the same or similar parameters.

**Tangent space projection of functional connectomes**

Functional connectomes (or FCs) are correlation matrices and hence lie on or inside the so-called Symmetric Positive Definite (SPD) manifold where the geometry is non-Euclidean[37]. The SPD cone of the correlation matrices is a Riemannian manifold. As mentioned in the Introduction section, vectorizing these correlation matrices directly and using these vectors as features is not ideal as the features are not independent due to the SPD constraint. Moreover, the canonical method of using Euclidean or correlation distance to compare vectorized correlation matrices is also sub-optimal because these matrices lie on an SPD cone. Therefore, a metric was introduced which accounts for the underlying non-Euclidean geometry of correlation matrices, called Affine-Invariance Riemannian Metric (AIRM)[38], or simply geodesic distance. This geodesic distance between FCs on the SPD manifold can be approximated by computing the Euclidean distance between tangent-FCs (FCs after tangent space projection).

Tangent space projection is a mapping technique that projects correlation matrices onto a tangent space that is Euclidean. The procedure is as follows. Correlation matrices are projected onto the tangent space relative to a selected reference point ($C_{ref}$) on the SPD cone. Such a reference point can be chosen in different ways[23] as detailed below (see Table 1). Once a reference point, matrix $C_{ref}$, has been chosen, a correlation matrix, $Q$, on the SPD manifold, can be projected using the following analytical formula:

$$\vec{Q} = log_m(C_{ref}^{-1/2} Q C_{ref}^{-1/2}) \qquad \text{(Equation. 1)}$$

where,

$\vec{Q}$ is the projected matrix on the tangent space as produced by $C_{ref}$

$Q$ is the SPD matrix on the manifold

$C_{ref}$ is the reference point/matrix on the manifold

$log_m$ is the matrix logarithm function.

Edges (matrix entries) in the tangent-space matrices are not inter-related, and thus no longer constrained by the SPD criterion. Hence, they can be vectorized. Furthermore, it has been shown[48] that the Euclidean distance in tangent space projections of any two correlation matrices approximates the underlaying Geodesic distance in the SPD manifold.

An important consideration for the tangent space projection of correlation matrices is that the projection requires a reference point on the manifold that should be *close* to all the correlation matrices[23]. This is the only point where the SPD manifold touches the tangent space. It should be noted that using a different reference point for each correlation matrix would result in each correlation matrix getting projected to a different tangent space. The reference matrix ($C_{ref}$) can be estimated in a number of different ways, most commonly by estimating a sample mean (or centroid) of the data, as shown in Table 1. Alternatively, a centroid can be estimated by using the identity matrix of the same dimension as the reference matrix.

Table 1: Equations for the estimation of five reference matrices ($C_{ref}$). $Q_i$ represents the $i^{th}$ correlation matrix in a set of correlation matrices (here functional connectomes).

| Reference Matrix ($C_{ref}$) | Equation |
|---|---|
| Euclidean | $Q_e = \frac{1}{N} \sum_i Q_i$ |
| Harmonic | $Q_h = \left(\frac{1}{N} \sum_i Q_i^{-1}\right)^{-1}$ |
| Log-Euclidean | $Q_{le} = Eaxp\left(\frac{1}{N} \sum_i Log(Q_i)\right)$ |

| Riemann | $Q_r = argmin\left(\sum_i d_G(Q_e Q_i)^2\right)$ |
|---------|--------------------------------------------------|
| Kullback | $Q_k = Q_e^{1/2}\left(Q_e^{-1/2} Q_h Q_e^{-1/2}\right)^\alpha Q_e^{1/2}$ |

As discussed above, for a given sample of FCs, computing their tangent-space projections requires two simple steps. First, one estimates a reference matrix ($C_{ref}$) from the sample FCs and then applies Equation 1 to project each FC onto the tangent space to obtain a tangent-FC.

To minimize biasing the ID rates of tangent-FCs, only one session of the FCs was used to estimate the $C_{ref}$ from a given sample. For example, for the test-retest dataset, only the test sessions were used to estimate $C_{ref}$ and subsequently project both test and retest FCs. Analogously, given a sample of MZ twins or of DZ twins, only one of the twins FCs was used to estimate $C_{ref}$.

**Distance metrics to compare functional connectomes**

The most commonly used estimates of similarity/dissimilarity between FCs[6,7] have been performed using Pearson's correlation coefficient, while other related approaches, such as Euclidean distance between the vectorized matrices[15], and the Manhattan ($L^1$) distance[49], have also been applied. However, these metrics are defined for vectors, not matrices so FCs and tangent-FCs need to be vectorized. Since FCs and tangent-FCs are symmetric matrices, vectorizing is applied to the upper triangular part of these matrices.

In this study, we used Pearson's correlation distance and Euclidean distance to compare vectorized FCs and vectorized tangent-FCs. We focused on correlation distance because Pearson's correlation (or equivalently correlation distance) is the most commonly used metric in the field of connectome fingerprinting[6,50]. As mentioned before, the Euclidean distance between tangent-FCs approximates Geodesic distance between the corresponding FCs on the SPD manifold and was therefore included as well.

Let $Q_1$ and $Q_2$ be two (square, symmetric) FC matrices of size $m \times m$, and $q_1 = vec(Q_1)$ and $q_2 = vec(Q_2)$ are the corresponding vectorized versions of size $m' = m(m-1)/2$. The mathematical formulae for correlation and Euclidean distance between these two matrices are described below.

*Pearson's Correlation Distance:*

Pearson's correlation between two vectors $q_1$ and $q_2$ is computed as:

$$r = \frac{\sum_{i=1}^{m'}(q_1(i) - \overline{q_1})(q_2(i) - \overline{q_2})}{\sigma_1 \sigma_2}$$

where

$r$ is the Pearson's correlation coefficient.

$q_1(i)$ and $q_2(i)$ are the $i^{th}$ elements of $q_1$ and $q_2$, respectively.

$\overline{q_1}$ and $\overline{q_2}$ are the sample means of $q_1$ and $q_2$, respectively.

$\sigma_1$ and $\sigma_2$ are the standard deviations of $q_1$ and $q_2$, respectively.

Finally, Pearson's correlation distance is simply defined as:

$$d_r = 1 - r$$

*Euclidean Distance:*

The Euclidean distance between two vectors $q_1$ and $q_2$ is computed as:

$$d_E = \sqrt{(q_1(1) - q_2(1))^2 + (q_1(2) - q_2(2))^2 + \cdots + (q_1(m') - q_2(m'))^2}$$

**Types of main-diagonal regularization**

The mathematical formulation of most of the reference matrices and tangent-space projection requires FCs to be invertible or full-rank (note the need to compute the inverse in Equation 1). When this is not the case, we can regularize these FCs by adding a scaled identity matrix, $\tau \times I$, which increases the value of their eigenvalues by $\tau$, ensuring that the matrices are invertible[1,8].

Canonically, a fixed regularization magnitude of $\tau = 1$ is used to achieve full-rank invertible matrices[8]. Alternatively, it has been recently shown that, for a given fMRI condition and granularity of parcellation, an *optimal* amount of regularization ($\tau^*$) can be estimated from a sample of FCs, by maximizing the test-retest ID rates based on geodesic distances[1].

Here, for tangent-FCs, we underwent the following steps to figure out the optimal amount of regularization that maximizes ID rates. For any given fMRI condition and parcellation granularity, the ID rates here were computed for a wide range of magnitudes for the regularization parameter, $\tau$. In particular:

$$\tau = \begin{cases} [0\ \ 0.01]\ and\ [0.5\ to\ 30, in\ steps\ of\ 0.5] & when\ all\ FCs\ are\ invertible \\ [0.01]\ and\ [0.5\ to\ 30, in\ steps\ of\ 0.5] & otherwise \end{cases}$$

An *optimal* regularization magnitude for a given fMRI condition and parcellation granularity can be computed using the steps described in Table 2.

**Table 2.** A step-by-step outline of how to estimate and apply an optimal regularization magnitude ($\tau^*$) to an FC dataset, such that individual fingerprint is maximized. This is a modified version of Table 4 of [1] for tangent-FCs.

| Step 1 | Estimate test and retest FCs per participant from the fMRI data |
|---|---|
| Step 2 | For a wide range of regularization magnitude ($\tau$): <br> a. Obtain a random sample of the FC dataset without replacement* <br> b. Regularize FCs by that regularization magnitude ($\tau$) <br> c. Project regularized FCs onto the tangent space using a specific reference matrix <br> d. Compute pairwise distances between tangent-FCs and obtain the identifiability matrix. <br> e. Estimate the ID rate from the identifiability matrix <br><br> *Random sampling without replacement is performed to estimate the mean and standard error of the ID rate with respect to the regularization parameters |
| Step 3 | Identify the optimal regularization magnitude ($\tau^*$), such that (mean) ID rate is maximized |

Since we are using Euclidean and correlation distances to compare tangent-FCs, we can estimate two different optimal $\tau^*$ for each given condition, granularity, and reference matrix:

1. $\tau^*_{Eud(\tan)}$: optimal regularization when Euclidean distance is used to compare tangent-FCs
2. $\tau^*_{corr(\tan)}$: optimal regularization when correlation distance is used to compare tangent-FCs

When comparing original FCs, only the upper triangular values of the matrices are used. Hence, main diagonal regularization would have no impact on the distances between FCs, whether using correlation or Euclidean distance. Geodesic distances between FCs would obviously be affected since the whole matrices are used to compute these distances. As for tangent-FCs, even though only the upper triangular values of matrices are used to compare them (using correlation or Euclidean distance), the whole matrices are used to projects FCs onto the tangent space (see Equation 1). Hence, distances between tangent-FCs are affected by the regularization introduced into the FCs.

## Results

In this work, we explored the effects of tangent space projection of FCs on the individual- and twin-fingerprinting. Identification (ID) rate[6] was used to quantify individual and twin fingerprints. For individual fingerprint, computation of ID rate required identifying an individual's FC from a population of FCs, given another FC of that individual. ID rate was simply the fraction of accurately identified individuals. When extending this concept to twin fingerprint, the process is analogous. That is, we try to identify a twin's FC from a population of FCs, given an FC of the corresponding twin. See Methods section *Quantification of fingerprinting* for details.

In particular, we investigated how ID rates are affected by the following factors: (i) *Choice of the Reference matrix $C_{ref}$*, (ii) *Main-diagonal Regularization* (i.e., weighted main diagonal regularization[1]), (iii) *Different fMRI conditions* (resting state and seven fMRI tasks), (iv) *Parcellation granularities from 100 to 900 brain regions (plus subcortical)*, (v) *Distance metrics*. Specifically, we used correlation and Euclidean distances to compare FCs as well as tangent-FCs. (vi) fMRI *Scan length* on resting state and when comparing task-based and resting state fingerprinting.

As an exploratory analysis, we first evaluated a simple scenario of a tangent space projection and its effect on ID rates of different fMRI conditions (Figure 2). Here we fixed the parcellation granularity to Schaefer 100 regions since this is the only granularity for which FCs for all fMRI conditions are full-rank and thus do not require any regularization before projection (see Methods section *Types of main-diagonal regularization* for details). For FCs, correlation distance (or correlation directly used as a similarity measurement) was the canonical metric[6,7]. As for tangent-FCs, we used the Euclidean distance between them on the tangent space that approximated the Geodesic distance on SPD manifold[48], as a geometry aware, more principled way of comparing FCs[1,8]. Thus, initially, we compared the ID rates of FCs using correlation distance ($ID_{corr(orig)}$), with that of the tangent-FCs using Euclidean distance ($ID_{Eud(tan)}$) (Figure 2). For comparison purposes, we also added the ID rates for FCs using Euclidean distance ($ID_{Eud(orig)}$). This exploratory analysis showed that tangent-FCs provided higher identification rates than both Euclidean and correlation distance FCs for all MRI conditions, with an increase in ID rates of 0.25-0.35. While this promising result cannot be extrapolated to other parcellations, it motivated further exploration of the effects of tangent space projections.

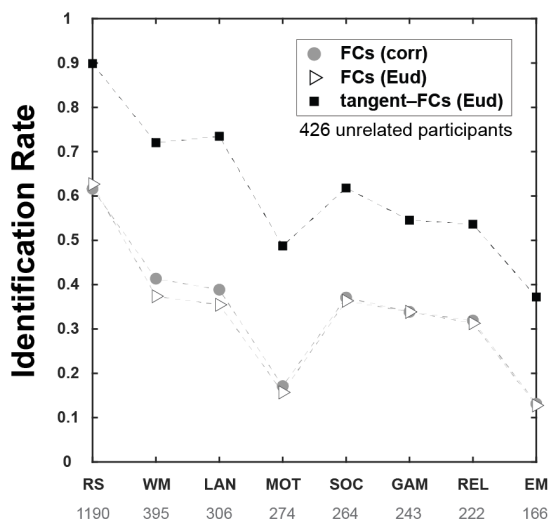

**Figure 2: Preliminary analysis on the effect of tangent-space projection on the ID rates using 426 unrelated participants.** Results are shown for all eight fMRI conditions (using entire scan length for each condition and session), Schaefer 100 parcellation granularity, and Riemann reference for $C_{ref}$. From left to right, the conditions are presented in descending order of scan length, as inscribed below the condition labels (in number of time-points). Since FCs for all conditions are full rank at 100 parcellation granularity, FCs were not regularized before projection onto the tangent space, i.e., $\tau = 0$. Legend indicates different scenarios: Light-grey circles represent ID rates using correlation distance ($ID_{corr(orig)}$), while hollow triangles represent Euclidean distance ID rates ($ID_{Eud(orig)}$). Black squares represent ID rates when Euclidean distance is used to compare tangent-FCs ($ID_{Eud(tan)}$). (Of note, the error bars reflecting the standard error of the mean across cross-validation resamples are small enough to be hidden by the symbols).

As the next step, we investigated higher parcellation granularities, where FCs for most of the conditions are rank-deficient (non-invertible), and hence cannot be projected as is. A fixed regularization magnitude of $\tau = 1$ was previously used to achieve full-rank invertible matrices[8]. Results with $\tau = 1$ regularization (which ensures full rank FCs for any granularity and any scan length) are shown in Figure 3. When using correlation distance and Euclidean distance on FCs, by increasing parcellation granularity (here evaluated up to Schaefer 900) we achieved higher ID rates for all fMRI conditions (Figure 3; Left panel). It is also worth noting that there are important differences in ID rates for the different conditions possibly due to the nature of the task and/or fMRI scan length (see analyses below to dissect the effect of the latter). Overall, ID rates obtained with correlation distance were higher or equal than those obtained for Euclidean distance across conditions and parcellations. Based on these results, from here onwards, we hence focused on the ID rates corresponding to the correlation-distance for FCs, i.e., $ID_{corr(orig)}$.

When assessing tangent-FCs with a fixed regularization of $\tau = 1$, the $ID_{Eud(tan)}$ curves are concave with respect to parcellation granularity, unlike $ID_{corr(orig)}$ and $ID_{Eud(orig)}$ that monotonically increase (Figure 3; right panel). While $ID_{Eud(tan)} > ID_{corr(orig)}$ for 100-300 parcellation granularity for all fMRI conditions, they decrease exponentially with increasing granularity, ultimately yielding low (below 30%) identification rates for all but resting-state fMRI conditions at the 900 parcellation granularity (Figure 3; right-panel). The results across fMRI conditions and parcellation granularities on tangent-FCs are qualitatively similar for all six $C_{ref}$ matrices.

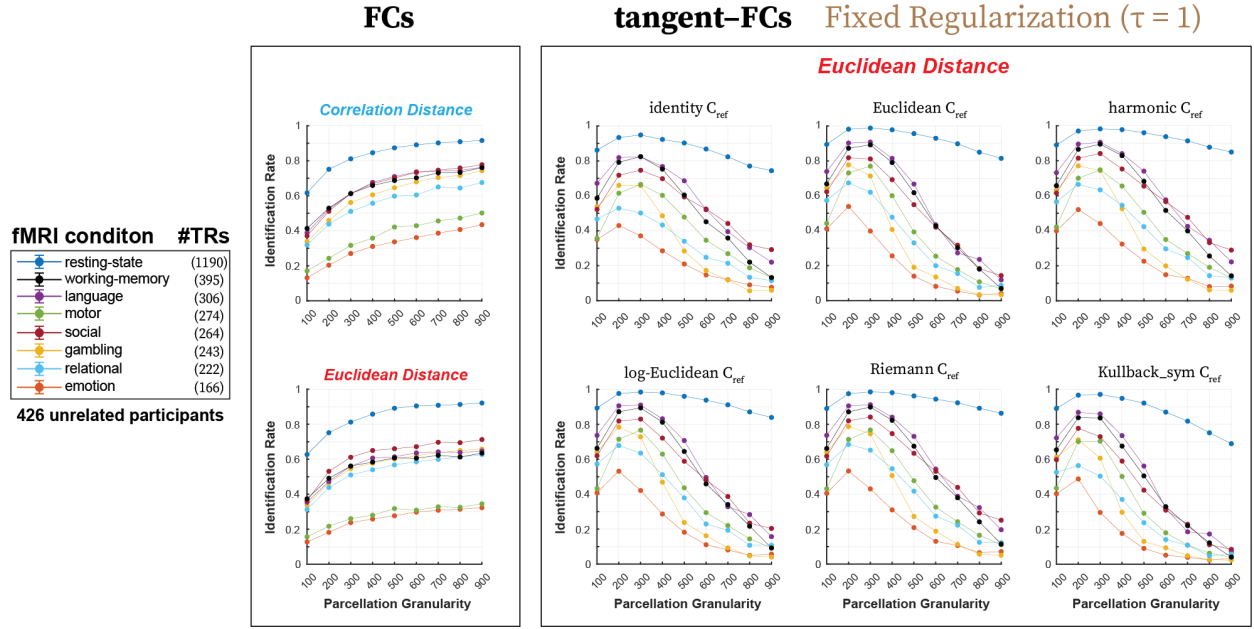

**Figure 3: Effect of tangent-space projection on ID rates using 426 unrelated participants when FCs are regularized ($\tau = 1$ for all cases) and Euclidean distance is used to compare tangent-FCs.** Results are shown for all eight fMRI conditions (utilizing maximum available TRs for each condition) and increasing granularity of Schaefer parcellations (100-900). Left panel shows ID rates for FCs using correlation and Euclidean distance metrics to compare FCs ($ID_{corr(orig)}$ and $ID_{Eud(orig)}$, respectively). Right panel shows ID rates for tangent-FCs when applying different reference matrices ($C_{ref}$). For tangent-FCs, Euclidean distance is used to compare FCs ($ID_{Eud(tan)}$). (Of note, the error bars reflecting the standard error of the mean across cross-validation resamples are small enough to be hidden by the symbols).

Using a fixed regularization ($\tau = 1$) led to low ID-rates at higher granularities in the tangent space. Hence, as explained in Methods section *Quantification of fingerprinting*, an optimal magnitude of regularization ($\tau^*$) that maximizes ID rate[1] was found for each fMRI condition and parcellation granularity. This procedure was originally proposed for original FCs when using Geodesic distance[1]. Instead, here we estimated an optimal $\tau^*$ using Euclidean distance on tangent-FCs, $\tau^*_{Eud(\tan)}$. This process was repeated for all fMRI conditions and parcellation granularities, allowing for different optimal $\tau^*$ across configurations. As already noted, regularization has no effect on the resulting ID rates for FCs when using correlation distance (see Methods section *Types of main-diagonal regularization* for details). Results for tangent-FCs using Euclidean distance showed that ID rates systematically increased for all conditions and parcellation granularities when using optimal regularization (Figure 4; right-panel). Importantly, $ID_{Eud(tan)} > ID_{corr(orig)}$ if optimal regularization $\tau^*_{Eud(\tan)}$ was applied, for all conditions and parcellation granularities. ID rates increased with increasing parcellation granularity, regardless of the reference matrix ($C_{ref}$). For resting state, $ID_{Eud(tan)}$ rates reached 100% accuracy for granularities of 400 and above. For all other conditions, $ID_{Eud(tan)}$ rates went above 80% at the parcellation granularity of 900, regardless of the reference matrix. This is in stark contrast to $ID_{corr(orig)}$. For instance, the highest $ID_{corr(orig)}$ rate achieved for 'emotion' condition was below 50%. Although the results are relatively consistent across different $C_{ref}$ matrices for tangent-FCs, the identity reference provided lower ID rates for lower granularities (100-200).

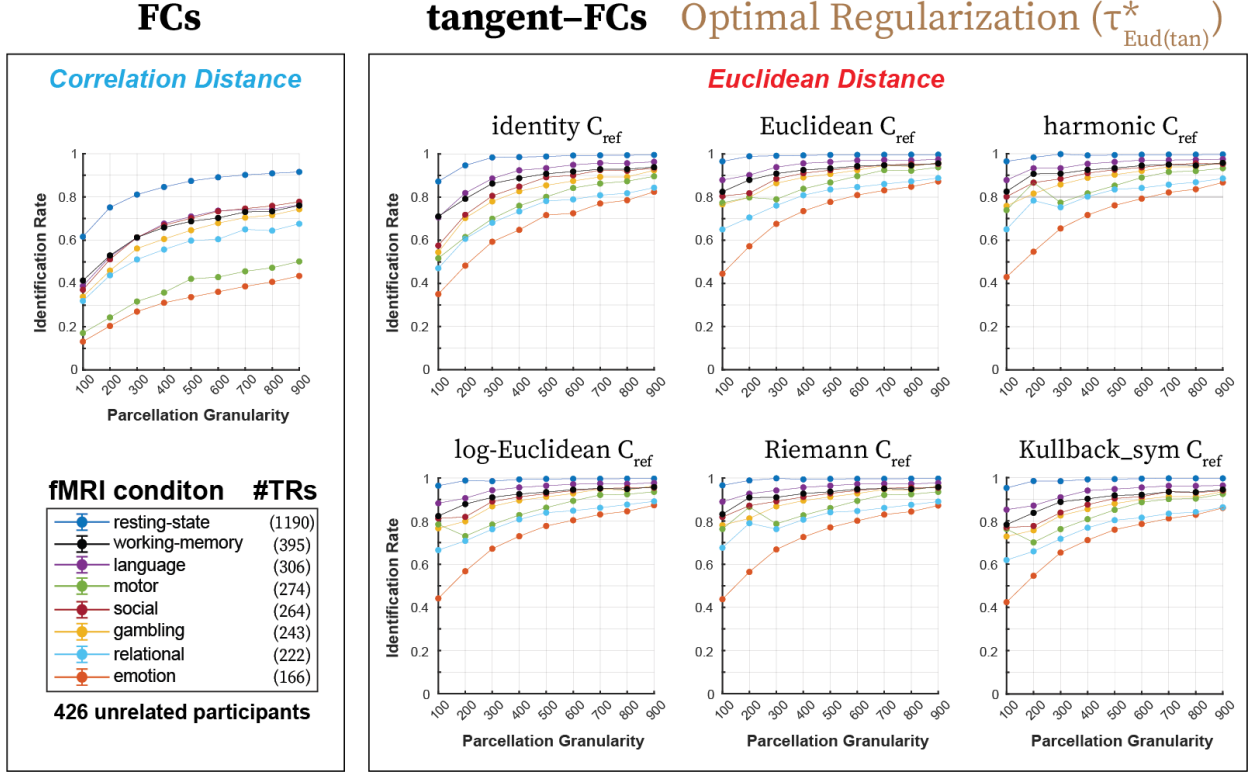

**Figure 4:** *Optimal Regularization ($\tau^*_{Eud(tan)}$) and Euclidean Distance* ─ **Effect of tangent-space projection on ID rates when FCs are regularized by *optimal* magnitude ($\tau^*_{Eud(tan)}$) and Euclidean distance is used to compare tangent-FCs.** Results are shown for all eight fMRI conditions (using entire fMRI scan length) and increasing granularity of Schaefer parcellations (100-900). For each fMRI condition and parcellation granularity, an *optimal* regularization magnitude was determined by the procedure in Table 2, and then the corresponding FCs were regularized by that magnitude. Left panel shows ID rates for FCs when correlation distance is used to compare FCs ($ID_{corr(orig)}$). Right panel shows the ID rates for tangent-FCs which are obtained by tangent-space projection of FCs using six different reference matrices ($C_{ref}$). For tangent-FCs, only Euclidean distance is used to compare FCs for this figure. (Of note, the error bars reflecting the standard error of the mean across cross-validation resamples are small enough to be hidden by the symbols).

Since correlation distance performed better than Euclidean distance in original FCs, we explored the ID rate performance when using correlation distance with tangent-FCs, i.e., $ID_{corr(tan)}$. Just as we estimated $\tau^*_{Eud(tan)}$ for each condition and granularity, analogously we estimated $\tau^*_{corr(tan)}$, i.e., the *optimal* regularization magnitude when correlation distance is used to compare tangent-FCs. Figure 5 shows the results for $ID_{corr(tan)}$ when FCs are regularized by $\tau^*_{corr(tan)}$ just prior to performing the tangent projection based on a reference matrix ($C_{ref}$). When compared to $ID_{Eud(tan)}$, $ID_{corr(tan)}$ rates were systematically higher for all conditions and granularities when the Riemann reference was used (see Figure 5). Remarkably, $ID_{corr(tan)}$ reached 100% for granularities above 300 for all conditions. $ID_{corr(tan)}$ rates for Riemann reference are closely followed by the $ID_{corr(tan)}$ rates with log-Euclidean and harmonic references, whereas Identity and Kullback references perform comparably to $ID_{Eud(tan)}$ (Figures 4 and 5).

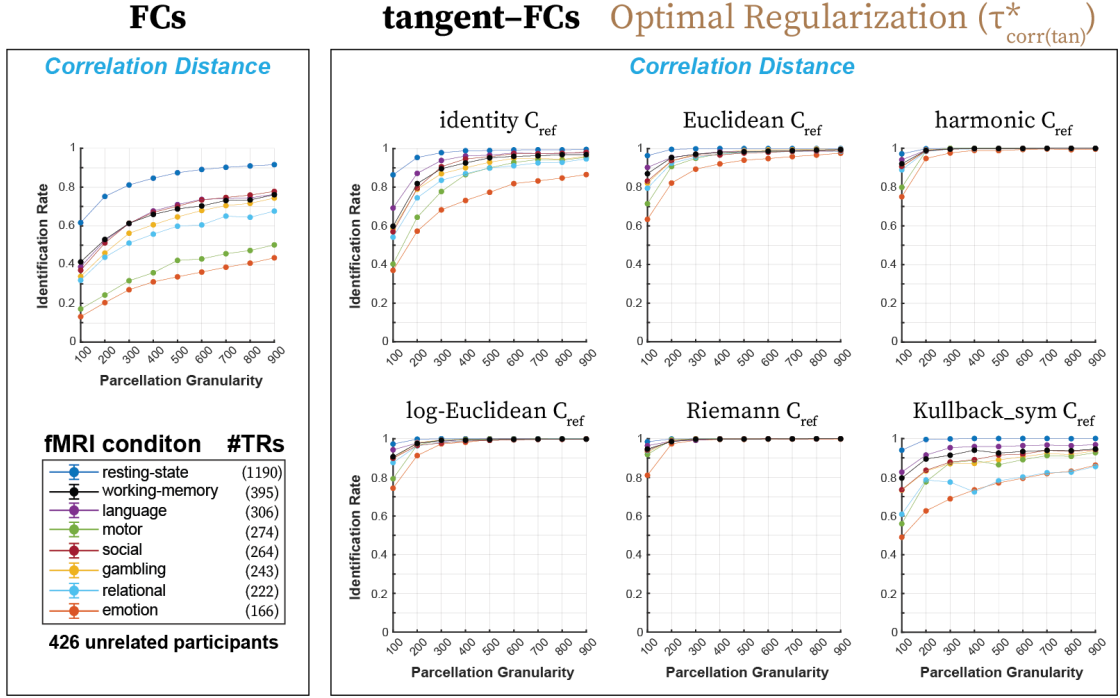

**Figure 5:** *Optimal Regularization ($\tau^*_{corr(tan)}$) and correlation Distance* — **Effect of tangent-space projection on ID rates when FCs are regularized by *optimal* magnitude ($\tau^*_{corr(tan)}$) and correlation distance is used to compare tangent-FCs.** Results are shown for all eight fMRI conditions (utilizing maximum available TRs for each condition) and increasing granularity of Schaefer parcellations (100-900). For each fMRI condition and parcellation granularity, an *optimal* regularization magnitude was determined by the procedure detailed in Table 2, and then the corresponding FCs were regularized by that magnitude. Left panel shows ID rates for FCs when correlation distance is used to compare FCs, i.e., $ID_{corr(orig)}$. Right panel shows the ID rates for tangent-FCs which are obtained by tangent-space projection of FCs using different reference matrices ($C_{ref}$). For tangent-FCs, only correlation distance is used to compare FCs for this figure, i.e., $ID_{corr(tan)}$. (Of note, the error bars reflecting the standard error of the mean across cross-validation resamples are small enough to be hidden by the symbols).

Together with assessing which distance, granularity, and condition maximizes ID rate, it is important to quantify the corresponding optimal levels of regularization required to reach such performance. Figure 6A summarizes the optimal regularization values that maximized ID rates for Euclidean distance ($\tau^*_{Eud(tan)}$) and for correlation distance ($\tau^*_{corr(tan)}$) as corresponding to the Riemann reference. It can be noted that $\tau^*_{Eud(tan)}$ is highly dependent on the condition and granularity, and for each condition, $\tau^*_{Eud(tan)}$ magnitudes increase with greater granularity. In contrast, $\tau^*_{corr(tan)}$ magnitudes are almost universally equal to the smallest non-zero regularization used, i.e. 0.01, except for resting state at granularity of 200 and 400-900, for which it is 0. A two-way ANOVA was performed to compare the effect of task and parcellation granularity on optimal regularization for both Euclidean and correlation distance. Results showed a significant (p<0.01) task effect and parcelation granularity effect on the optimal regularization associated to Euclidean distances and no significant associations for correlation distance. Note that we did not include resting state for this analysis because the optimal results obtained ($\tau^*_{corr(tan)}=0$) are not possible for any other tasks for all parcellation granularities (see Methods section *Types of main-diagonal regularization*).

Based on optimal regularization values $\tau^*_{Eud(tan)}$ and $\tau^*_{corr(tan)}$, the corresponding ID rates are shown in Figure 6B (top left and bottom right, for Euclidean distance and correlation distance respectively). To explore how $ID_{corr(tan)}$ and $ID_{Eud(tan)}$ are sensitive to suboptimal regularization, we obtained both ID rates when using the optimal regularization of each other. Specifically, we obtained ID rates of tangent-FCs regularized by $\tau^*_{corr(\tan)}$ when using Euclidean distance (Figure 6B top right), as well as FCs regularized by $\tau^*_{Eud(\tan)}$ when using correlation distance (Figure 6B bottom left). It is noteworthy that $ID_{Eud(\tan)}$ are severely affected by $\tau^*_{corr(\tan)}$, with most rates dropping below 10%. On the other hand, $ID_{corr(\tan)}$ remained nearly invariant to $\tau^*_{Eud(\tan)}$, reaching an almost perfect 100% identification rate for most of the configurations. To show these contrasting behaviors with respect to regularization, the ID rate differences (ID rate gain) between using optimal and suboptimal regularization are shown in Figure 6C.

Based on the results shown in Figures 2−6, we can summarize a few key findings: (1) tangent-FCs have higher ID rates than FCs, (2) Riemann reference is the best choice for a reference matrix yielding the highest ID rates, (3) for both FCs and tangent-FCs, correlation distance provides higher ID rates than Euclidean. (4) tangent-FCs require a miminal and almost universal (across conditions and granularities) regularization to achieve maximal ID rates. (5) When using Euclidean distance, ID rates are very sensitive to regularization, but when using correlation distance ID rates are barely affected.

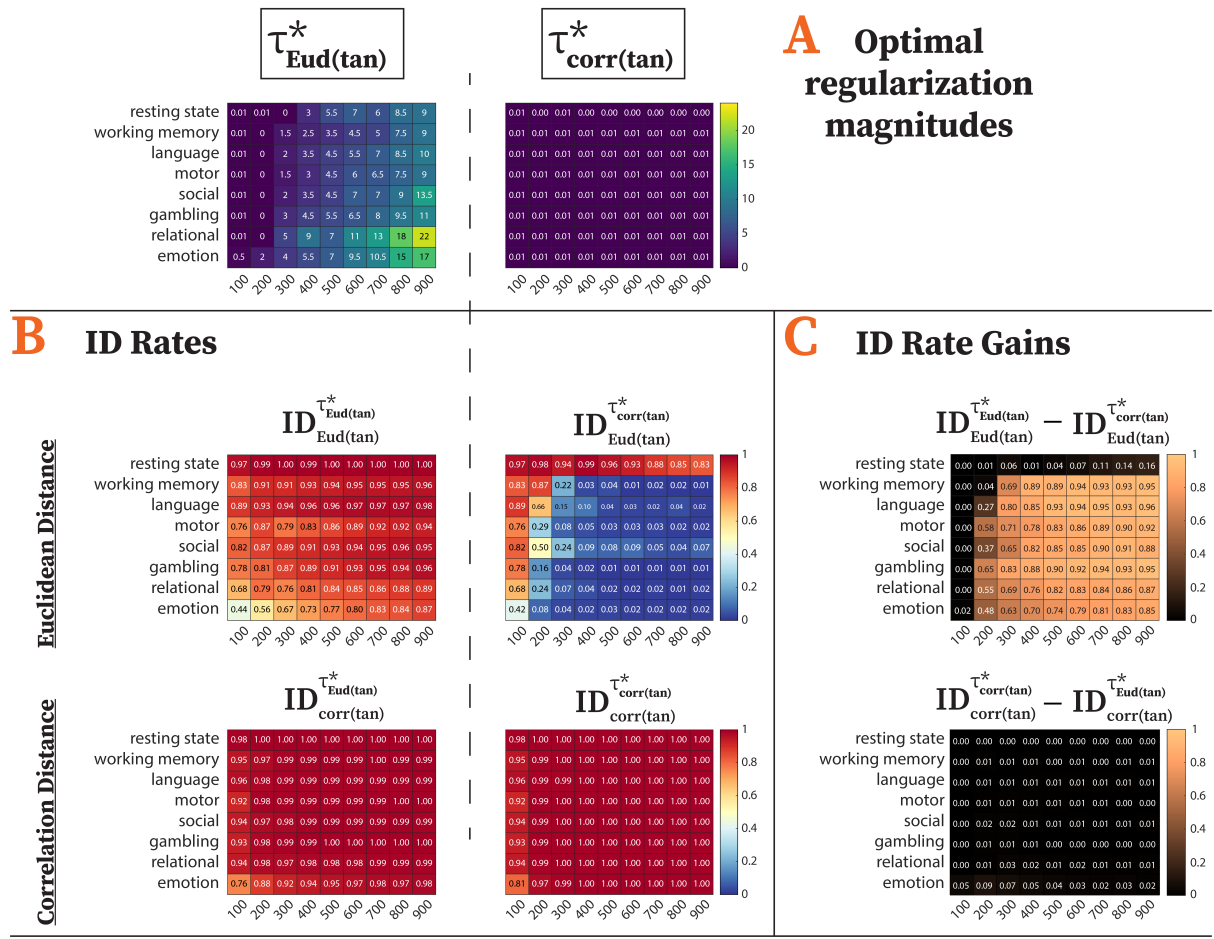

**Figure 6: Effect of optimal regularization on ID rates.** Only results corresponding to the Riemann reference are presented. (A) optimal regularization magnitudes for all the fMRI conditions and parcellation granularities when Euclidean ($\tau^*_{Eud(\tan)}$; left) and correlation ($\tau^*_{corr(\tan)}$; right) distance is used to compare tangent-FCs (left). (B) ID rates corresponding to the optimal regularization

magnitudes shown in (A). The subscript in each title indicates the distance metric used to compare the FCs: Eud(tan) for Euclidean and corr(tan) for correlation distance on tangent-FCs. The superscript indicates the type of optimal regularization that was used to regularize FCs. (C) ID rate gains when optimizing regularization for each distance: element-wise difference in the ID rates shown in (B) within Euclidean and correlation distance. The title at the top of each matrix shows this difference in an equation form.

In order to take a step further beyond test-retest in fingerprinting analysis of functional connectivity and in assessing the impact of the Riemannian operations shown above, we extended our analyses to monozygotic (MZ) and dizygotic (DZ) twin fingerprinting (see Methods section *Tangent space projection of FCs*). This builds and expands on a former study looking at tangent FCs inter-subject variability on twins and siblings[22]. Here we hypothesized an increasing fingerprint in the FCs and tangent-FCs, as follows: $ID_{test-retest} > ID_{MZ} > ID_{DZ}$. This "Fingerprint Gradient" reflects the genetic and shared environment gradients. Based on the fingerprinting results obtained for test-retest, we focused on the Riemann reference matrix while exploring all conditions and granularities.

Figure 7 shows results for $ID_{TR}$, $ID_{MZ}$ and $ID_{DZ}$ when using correlation distance on FCs (Figure 7, top row), Euclidean distance on tangent-FCs (Figure 7, middle row) and correlation distance on tangent-FCs (Figure 7, bottom row). We capped each cohort to 63 pairs, matching the smallest cohort size (DZ). The Fingerprint Gradient is present in both FCs and tangent-FCs. Across all fMRI conditions, parcellation granularities, and the three cohorts, ID rates were systematically higher for tangent-FCs than for FCs. Furthermore, for tangent-FCs, $ID_{corr(\text{tan})} \gg ID_{Eud(\text{tan})}$ across all scenarios, except for DZ twins at Schaefer-100 granularity where relational and emotion conditions had comparable ID rates. Results along parcellation granularity showed high variability in $ID_{corr(\text{tan})}$ for DZ, with parcellations 400 to 600 showing large fluctuations.

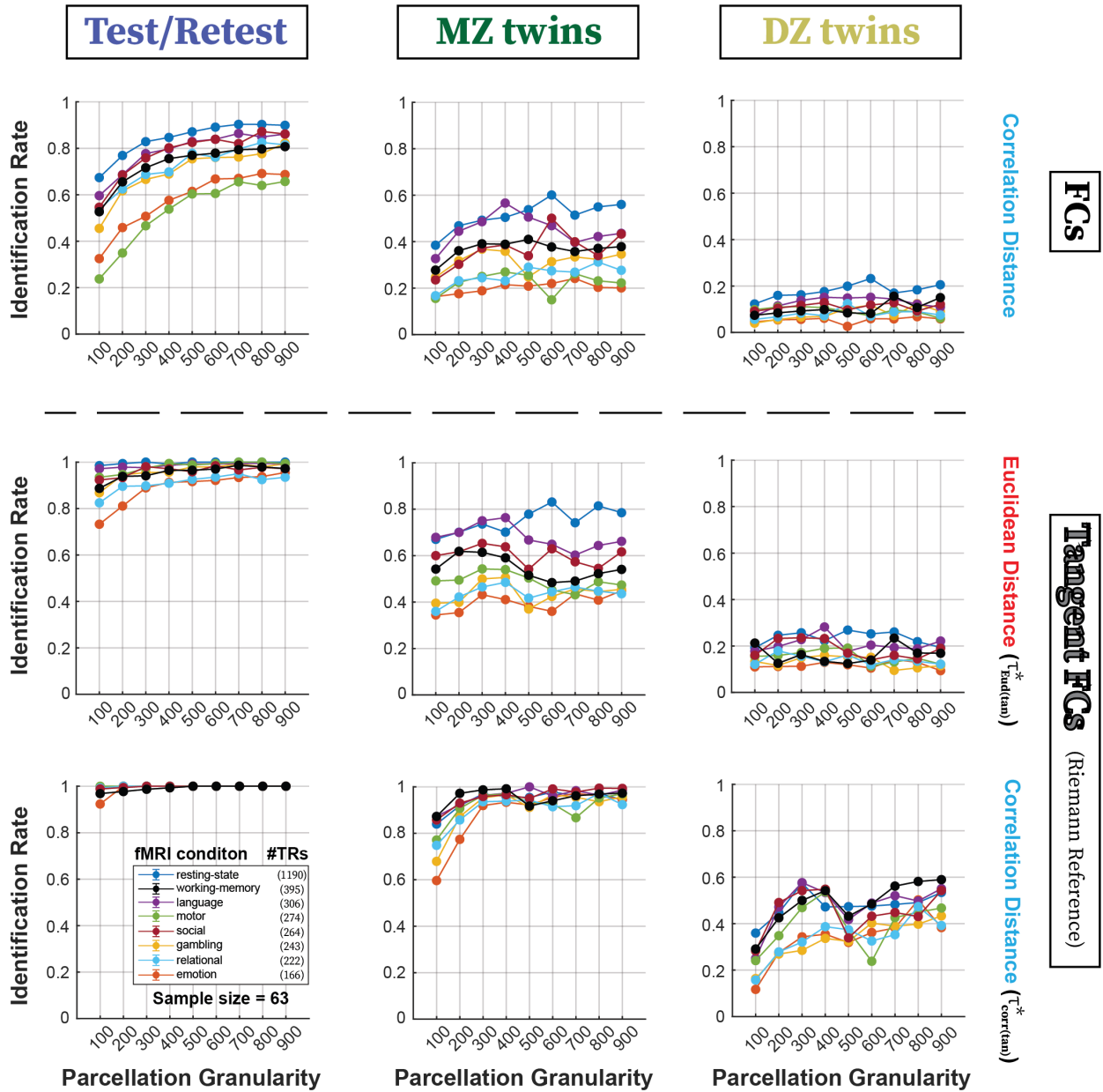

**Figure 7: Effect of tangent-space projection and distance metric (correlation, Euclidean) on the Fingerprint Gradient.** Results are shown for all eight fMRI conditions (using maximum available TRs for each condition) and increasing granularity of Schaefer parcellations (100-900). Top row shows ID rates for FCs when correlation distance is used to compare FCs, i.e., $ID_{corr(orig)}$. The ID rates for tangent-FCs using the Riemann reference are shown when using Euclidean distance (middle row) and correlation distance (bottom row). The corresponding optimal regularization values ensure that maximum available ID rates are presented for each given scenario. Sample sizes across the three cohorts (Test/Retest, MZ, and DZ twins; sample size = 63 pairs) were matched before computation of ID rates to enable meaningful comparisons. (Of note, the error bars reflecting the standard error of the mean across cross-validation resamples are small enough to be hidden by the symbols).

We also assessed the effects of fMRI scan length (number of TRs) on ID rates for the resting state with and without tangent space projections, and an intermediate granularity (Schaefer-400; Figure 8). Note that other granularities produce similar results (not shown). Overall, ID rates increase with increasing number of TRs for all three cohorts, and for FCs and tangent-FCs. This is consistent with previous findings of higher ID rates associated with longer fMRI scan lengths[1,6–8]. Also, analogously to the results observed in Figures 4 and 5, we noted that $ID_{corr(tan)} > ID_{corr(orig)}$, and $ID_{corr(tan)} > ID_{Eud(tan)}$. In addition, $ID_{corr(tan)}$ reached maximal values using far fewer TRs than $ID_{corr(orig)}$ or $ID_{Eud(tan)}$.

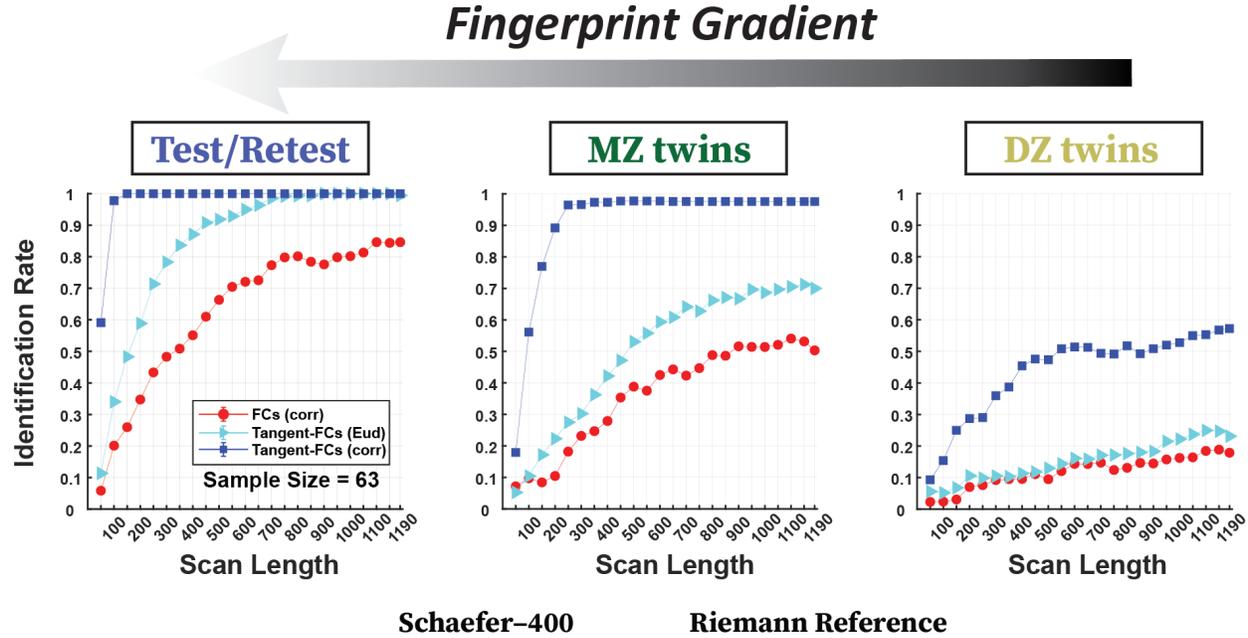

**Figure 8: Effect of resting-state fMRI scan length (number of TRs) on the Fingerprint Gradient, and interaction with tangent space projections and the distance metrics (correlation, Euclidean).** Results are shown for parcellation granularity of 400, and Riemann reference for projecting FCs into a tangent space. Left panel shows the ID rates for FCs and tangent-FCs for unrelated participants when the fMRI scan length increases (x-axis shows the number of TRs used to construct FCs). Middle and right panels show results for the MZ and the DZ twins respectively. The corresponding optimal regularization values are used to ensure that maximum available ID rates are presented for each given scenario. (Of note, the error bars reflecting the standard error of the mean across cross-validation resamples are small enough to be hidden by the symbols).

Results in Figure 8 highlight the important effect of fMRI scan length in ID rates on resting-state functional connectivity. Figure 9 compares task and resting state ID rates when accounting for different scan lengths across fMRI conditions. For all given scenarios, when matching fMRI scan length of rest to each task, task conditions had higher ID rates than resting state (except when perfect ID rate is reached by both). This is consistent with previous findings where Geodesic distance was used to compare FCs[8]. In addition, this trend is observable not only in test-retest of unrelated participants, but across the entire Fingerprint Gradient which includes MZ and DZ participants (Figure 9, middle and right).

All results are thus far derived from the HCP young-adult dataset. To validate the identifiability of tangent-FCs using correlation distance, we compared $ID_{corr(tan)}$ and $ID_{corr(orig)}$ for a separate dataset (181 healthy participants) collected at a different site (see dataset details in Methods section *Validation dataset*). Figure 10 shows that in this cohort, $ID_{corr(tan)}$ achieves 100% across all parcellation granularities (Schaefer 100-1000) while $ID_{corr(orig)}$ rises gradually from below 60% to above 90%. Notice that HCP young-adult dataset and validation dataset are not only acquired with different fMRI sequences but treated with different preprocessing pipelines. Despite these differences, tangent-FCs under a configuration learned on the HCP young-adult outperform original FCs in terms of identifiability when correlation distance is used, and again, provides results invariant to parcellation granularity.

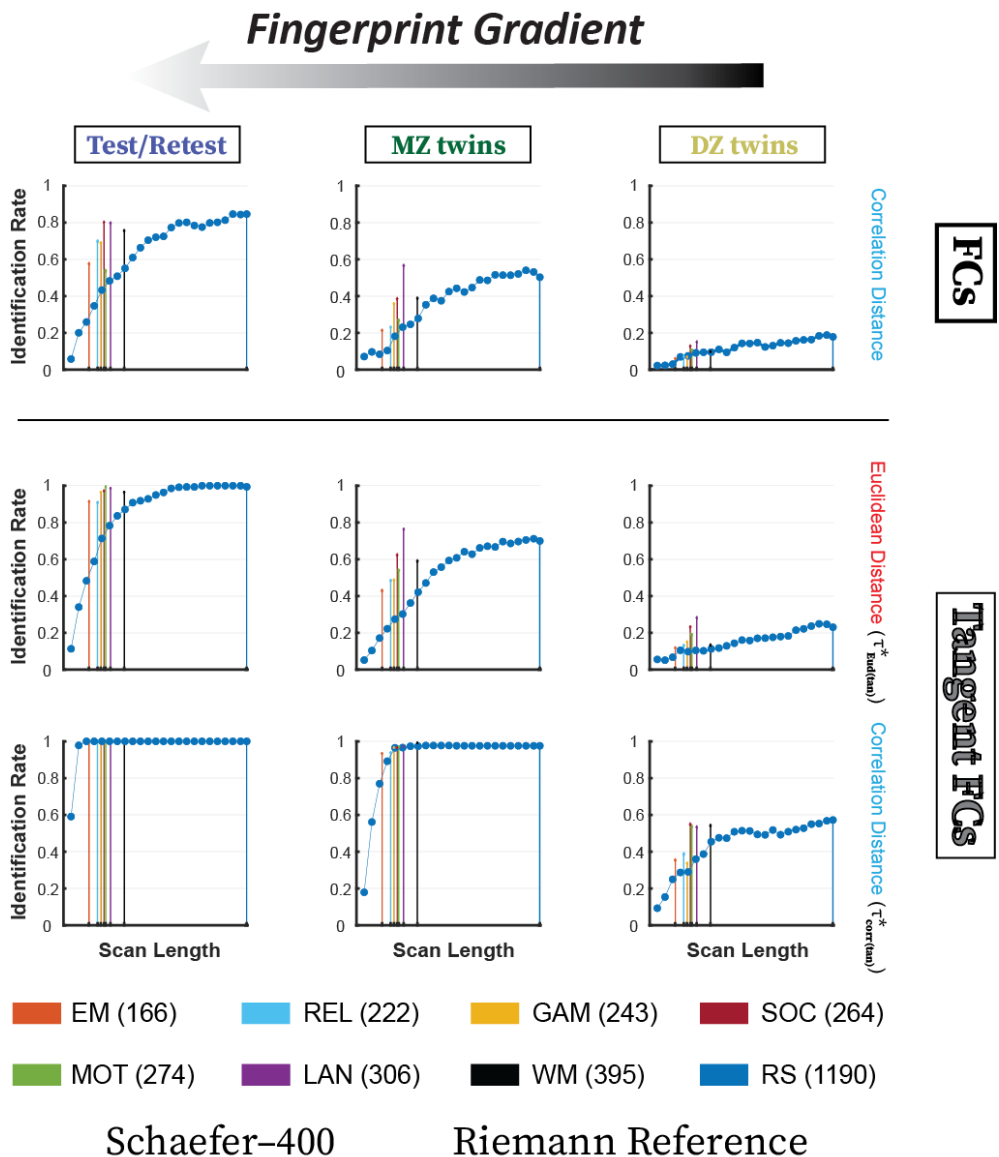

Figure 9: **Effect of the fMRI scan length on the Fingerprint Gradient for resting state *vs* task conditions.** Blue-curve shows the ID rates for resting state data with the fMRI scan length trimmed to shorter and longer than the task conditions (50 to 1190 TRs in steps of 50). Results are shown only for the parcellation granularity of 400, and when Riemann reference is used to project FCs into the tangent space. Left, middle, and the right columns show results for the unrelated test-retest participants, MZ twins, and the DZ twins, respectively. Top row shows results for the FCs when correlation distance is used ($ID_{corr(orig)}$), and bottom rows show the results for tangent-FCs when Euclidean ($ID_{Eud(tan)}$; second row) and correlation ($ID_{corr(tan)}$; third row) are used. Sample size (number of FCs) was matched across the three groups according to the smallest sample size (63 pairs). For tangent-FCs, when Euclidean distance was used, FCs were regularized by optimal magnitude $\tau^*_{Eud(tan)}$, whereas when correlation distance was used, FCs were regularized by $\tau^*_{corr(tan)}$. This ensured maximum available ID rates for each given scenario. (Of note, the error bars reflecting the standard error of the mean across cross-validation resamples are small enough to be hidden by the symbols).

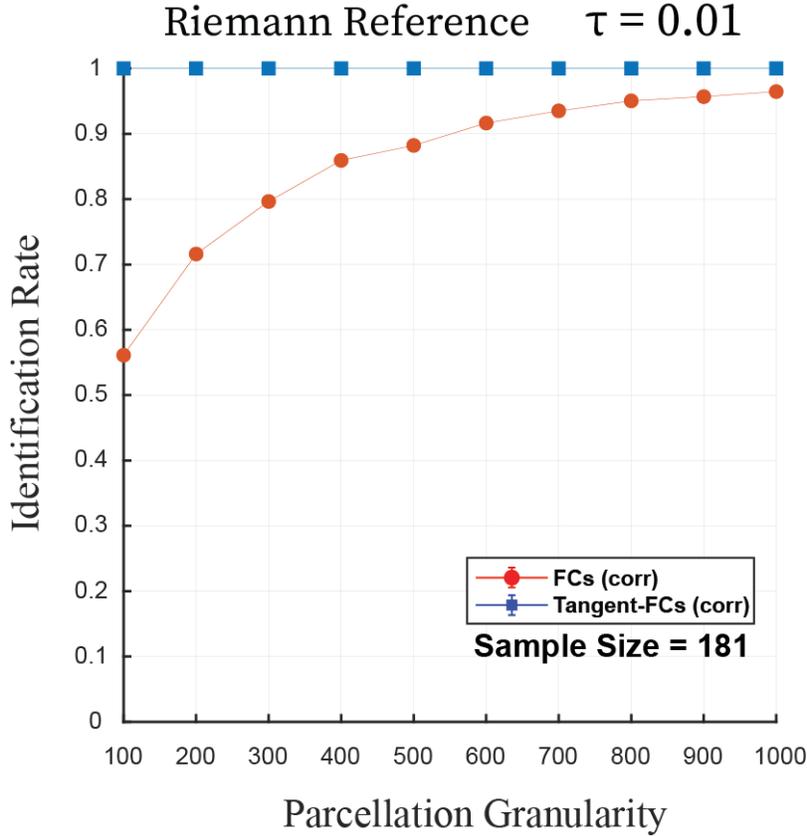

**Figure 10: Validation dataset. Effect of parcellation granularity and tangent-space projection on ID rates on a cohort of 181 healthy controls.** Results are shown for resting state condition and increasing granularity of Schaefer parcellations (100-1000). ID rates for FCs and tangent-FCs are shown when correlation distance is used to compare FCs, i.e., $ID_{corr(orig)}$ and $ID_{corr(tan)}$. For each parcellation granularity, a fixed regularization magnitude 0.01 and Riemann reference are used in tangent-space projection of FCs. This configuration is based on the results obtained for the HCP young-adult. (Of note, the error bars reflecting the standard error of the mean across cross-validation resamples are small enough to be hidden by the symbols).

## Discussion

Our starting point was that FCs, as correlation matrices, are part of the SPD manifold and hence distances between FCs are better measured along the geodesics on the manifold. Using geodesic distance has been proven to lead to higher identification rates[1,8]. While geodesic distance provides a more principled distance criterion for FC comparisons, it does not truly transform the FCs for further analysis. In other words, any further analysis on functional edges of FCs will remain bounded by the SPD criterion. This is why tangent space projections are relevant. First, the geodesic distance between original FCs on the manifold can be approximated by the Euclidean distance between tangent-FCs. Second, elements in tangent-FCs are not inter-related anymore and can be analyzed as independent features. Lastly, tangent-FCs have been proven to be better predictors of diseases and aging compared to the original FCs[19,21,24–26]. With the aforementioned advantages, no study has assessed the impact of tangent-FCs in fingerprinting. To fill such gap, we hypothesized that tangent-FCs have better fingerprinting than original FCs and this hypothesis is proved in this manuscript.

Specifically, we explored how the tangent space projection of FCs using Riemannian Geometry affects individual and twin fingerprint in FCs. In particular, we exhaustively explored how ID rates are affected by six factors. We found that: (i) Riemann as well as log-Euclidean were the matrix references that systematically leading to higher ID rates for all configurations assessed. (ii) In tangent-FCs, *Main-diagonal regularization prior to tangent space projection* was critical for ID rate when using Euclidean distance, whereas regularization barely affected ID rates when using correlation distance. (iii) When *evaluating different fMRI conditions*, it was found that ID rates were dependent on condition and fMRI scan length. (iv) *Parcellation granularities* were key for ID rates in FCs, as well as in tangent-FCs with fixed regularization. Optimal regularization of tangent-FCs when using either Euclidean distance or correlation distance mostly removed such effects. (v) *Correlation distance in tangent-FCs* outperformed any other distance metrics on FCs or on tangent-FCs across the entire fingerprint gradient. (vi) ID rates tended to be higher in tasks relative to resting state when accounting for fMRI *scan length*.

In the next subsections we further discuss these results, together with limitations and further work.

**Effect of reference matrix ($C_{ref}$) in fingerprinting on tangent-FCs.** A potentially important factor to consider when performing tangent space projections of FCs is the choice of $C_{ref}$[23]. Our results indicate that different choices (see Table 1 for different $C_{ref}$ evaluated) led to very similar ID rates when using Euclidean Distance with fixed regularization (Figure 3). Analogously, when using Euclidean distance with optimal regularization, very similar ID rates were achieved across different $C_{ref}$ (see Figure 4). However, when using correlation distance, we showed that Riemann reference led to the highest $ID_{corr(tan)}$ rates, followed closely by log-Euclidean, and harmonic (Figure 5). Overall, Riemann, as well as log-Euclidean references seem to be robust choices regardless of fMRI condition, parcellation granularity, distance used, and regularization.

**Effect of weighted main-diagonal regularization in fingerprinting on tangent-FCs**. In tangent-FCs, *main-diagonal regularization* was critical for ID rate when using Euclidean distance, whereas ID rates were barely affected when using correlation distance (see Figure 6). In other words, ID rates based on correlation distance seemed to be invariant to regularization, whereas ID rates based on Euclidean distance were highly sensitive to it. Hence, when using correlation distance, minimal regularization (such as 0.01) appears robust and ensures the invertibility of FCs while minimally changing the original matrix. Our results indicate that this is the case for any fMRI condition and parcellation granularity. Note that such a finding also avoids the computational time required to find an optimal regularization for each condition and/or granularity on each dataset.

**Effect of *different distance metrics in fingerprinting*.** Euclidean and correlation distance were used to compare FCs as well as tangent-FCs. For FCs, correlation distance contributes to a small improvement in ID rates compared to Euclidean distance. However, for tangent-FCs, using correlation distance systematically improved ID rates across fMRI conditions, parcellation granularities and the fingerprint gradient. In addition, tangent-FCs comparisons using correlation distance required a small fraction (around 150 resting state volumes for unrelated test-retest participants, 250 volumes for MZ and 600 volumes for DZ) of the fMRI scan length to achieve maximal reliability across the fingerprint. Such result was not achieved when using Euclidean distance.

Euclidean distance in tangent-FCs is supported mathematically as it approximates the underlying Geodesic distance in the SPD manifold[37]. However, there is a lack of literature on using correlation distance in tangent space projections of correlation or covariance matrices. The improvements in fingerprinting observed here with correlation distance as compared to Euclidean distance in tangent-FCs could be related to the "curse of dimensionality" phenomenon[58]. Future work posed on this finding would better characterize the origin of these differences.

**Effect of parcellation granularities in fingerprinting.** More fine-grained parcellations resulted in higher ID rates for FCs as well as for tangent-FCs when using Euclidean distance and optimal regularization. When assessing tangent-FCs with optimal regularization and correlation distance, the contribution of parcellation granularity to ID rates was very small, since perfect ID rates were achieved as low as 200-300 parcels in most cases. This trend was less apparent for twins, where more fine-grained parcellations did not contribute greatly to ID rates and indeed some fluctuations are observed.

Higher granularity for parcellation in a prediction algorithm increases prediction accuracy, but at the cost of poorer feature weight reliability[36]. Based on the observation above, we argue that by using the right distance metric (i.e., correlation distance) and a suitable tangent space (Riemann reference with minimal regularization), the best of both worlds (high prediction accuracy, high feature weight reliability) is possible.

**Tangent-FCs are more reliable phenotypes than FCs**

Functional connectomes (FCs) are correlation matrices that lie on or inside the Symmetric Positive Definite (SPD) manifold. Since the geometry in the SPD manifold is non-Euclidean[37], using either Euclidean distance or correlation distance to compare FCs is suboptimal[1,8]. Importantly, entries in tangent-FCs are not inter-related[19,59] and hence can be vectorized and compared using Euclidean distance and correlation distance. So, we hypothesized that tangent-FCs would lead to an enhanced reliability as compared to FCs.

Overall, we showed that tangent-FCs have higher ID rates than FCs (Figure 2−6). When no regularization is required to project FCs onto the tangent-space, ID rates are much higher than those obtained for FCs (Figure 2). When regularization is necessary, an optimal amount of regularization combined with tangent projection, led to considerably higher ID rates than FCs (Figure 4−6).

The only situation where tangent-space projections led to lower ID rates was when a fixed amount of regularization and Euclidean distance was used (Figure 3). For instance, this is the case when using fixed regularization of $\tau = 1$ and Euclidean distance with tangent-FCs, which has been the canonical choice so far[1,48,60]. Our results indicate that this choice may lead to systematically lower ID rates. This considerable reduction in reliability in tangent-FCs would in turn affect the performance of any classification, prediction, or inference algorithm that uses tangent-FCs as inputs. This is especially true for higher parcellation granularities (Figure 3).

As reliable connectivity objects that show high fingerprinting, tangent-FCs have great potential applications to fingerprinting and assessment of disease progression. For fingerprinting, here we simply computed pairwise distances between matrices (based on Euclidean distance and correlation distance) and measured identification rates relying on a relatively simple nearest neighbor approach. In scenarios where the only purpose is to maximize fingerprinting, supervised mapping methods like support vector machines or linear discriminant analysis might outperform the fingerprinting results presented in this work when using tangent-FCs (specially for DZ twins or for short scanning length). However, this does not undermine the significance of tangent-FCs carrying a much higher fingerprint than FCs, regardless of the classifier nature. When studying disease progression, the entries of tangent-FCs are not inter-related and can be examined individually[17,19]. In fact, tangent-FCs have been successfully applied to cognition and in predictions of disease progression[19,21,24,25]. Our results of tangent-FCs being more reliable phenotypes than regular FCs lay the groundwork for those applications, illustrate why tangent-FCs seem to be better predictors than FCs, and motivate future work of tangent-FC applications to developing disease biomarkers. This should include assessing functional connectivity in other species and possibly in other than fMRI modalities.

**Optimal recipe for fingerprinting using tangent-FCs.**

Despite the expected heterogeneity in ID rates across different datasets, some of our results seem generalizable. We aimed to find a set of *optimal* parameters that would uncover fingerprints by finding the best projection of the FCs onto the tangent space where inter-individual differentiability is maximized. Based on our data, *we recommend regularizing FCs by a small non-zero magnitude (say 0.01), doing tangent space projection of FCs using Riemann reference matrix and then comparing tangent-FCs with correlation distance.* Note that we tested this recipe in a separate validation dataset where both fMRI acquisition sequence and preprocessing pipeline were different from HCP young-adult dataset. The ID rates achieved 100% across all parcellation granularities. Following the optimal recipe, these tangent-FCs, as a phenotype, are potentially highly reliable, which makes them useful translationally in the monitoring and/or subsequent treatment of cognitive and behavioral disorders.

Correlation distance in tangent-FCs was practically invariant to regularization, although it does not follow that future predictors or biomarkers would also be invariant to regularization. The relationship between predictable and reliable functional connectomes has been recently debated[4,61] and presented as a possible dichotomy[61]. Our results do not support such a dichotomy. Instead, the results indicate that tangent space projections, previously reported as better predictors of disease, cognition and behavior than regular FCs[19,21–23], yield FCs with higher fingerprinting not only in the test-retest samples but also along the fingerprint gradient. It does so for all parcellation granularities and fMRI conditions evaluated.

As we outlined, tangent-FCs result in higher fingerprint and are better predictors because their functional edges are not bounded by the SPD criterion and are therefore a set of independent measurements. Many classifier algorithms benefit from avoiding inter-related features, and our results show that identification rates benefit as well. Inter-related features limit not only the performance but more importantly, generalizability of machine learning algorithms, resulting in a lack of reliable and robust clinical biomarkers for brain disorders using brain connectomic data. This limitation can be addressed by projecting FCs from the SPD manifold onto an "ideal" tangent space (tangent-FCs), which is Euclidean and hence allows the use of Euclidean algebra and calculus.

**Effect of fMRI scan length and fMRI conditions in fingerprinting.**

Scan length also affected ID rates in resting state connectivity. Tangent-FCs with correlation distance clearly outperformed tangent-FCs with Euclidean distance and FCs with correlation distance. ID rates reached for test-retest, MZ and DZ for tangent-FCs with correlation distance at 250 scan volumes were unachievable for the other two configurations even when using the entire scan (1190 volumes). The only exception was tangent-FCs with Euclidean distance for test-retest, which achieved the same ID rate after 900 volumes. Overall, the gain in ID rates from tangent-FCs with correlation distance cannot be compensated by simply extending the fMRI scan length, which is also not practical in clinical populations.

When evaluating different fMRI conditions, ID rates tended to be higher in tasks with respect to resting-state connectivity after accounting for the fMRI scan length for test-retest MZ and DZ (Figure 9). For instance, we showed that when matching fMRI scan length, language, working-memory, social and emotion conditions have much higher ID rates across the Fingerprint Gradient, consistent with previous findings where Geodesic distance was used to compare FCs. The only exception happened when tasks and rest achieve perfect ID rate (specifically test-retest and MZ cohorts when using correlation distance on tangent-FCs).

**Fingerprint Gradient: a more comprehensive metric of phenotypic reliability**

The ID rate metric[6] is used as a measure of the amount of fingerprint in a dataset. In turn, a higher fingerprint is reflective of higher phenotypic reliability. Previously, fingerprinting has been estimated using test-retest reliability[1,6–8,62]. In this work, we extended that concept to include twin-fingerprints (MZ and DZ) and proposed the 'Fingerprint Gradient' as a more comprehensive measure of phenotypic reliability. Such gradient relies on expected identifiability based on shared characteristics, with the same person measured twice being thought of having highest chance of sameness (followed by MZ and then MZ twins). We hypothesized that a phenotype with a higher test-retest fingerprint would also have a higher twin-fingerprint. This hypothesis was based on the framework of shared genetics and environment: test-retest FCs of an individual should be the most similar as they obviously share 100% of genetics and environment; MZ twins should follow as they share the same genetics, but the shared environment is likely to be high but not complete. Finally, DZ twins should be the least similar to each other as they share ~50% of the genetics, and the shared environment is <100%, like MZ twins. This hypothesis was shown to be true when evaluated in the HCP dataset, as tangent-FCs have higher fingerprints than FCs across the Fingerprint Gradient (Figure 7−9). Thus, in the future, when a new framework or a phenotype is to be tested in the connectomics, we recommend the use of the Fingerprint Gradient as a metric whenever possible, instead of mere test-retest fingerprint.

Interestingly, the ID rate pattern along scan-length on tangent-FCs with Euclidean distance for test-retest is very similar to tangent-FCs with correlation distance for MZ. Analogously, the ID rate pattern along scan-length on tangent-FCs with Euclidean distance for MZ is very similar to tangent-FCs with correlation distance for DZ. Overall, replacing Euclidean distance by correlation distance practically "moves up" one step in the Fingerprint Gradient in terms of ID rates.

Additionally, as shown in Figure 8, ID rates from Euclidean on tangent-FCs outperform ID rates from correlation on FCs for test-retest and MZ, but not for DZ. This may suggest that Euclidean distance on tangent-FCs is not able to uncover additional fingerprints (with respect to FCs) when genetics are not the same.

**Limitations and further work**

Our study has several limitations. From a theoretical standpoint, we lack of a mathematical demonstration of why correlation distance systematically outperformed Euclidean distance. Our empirical results suggest further exploration of the geometric role of correlation distance in tangent space projections, for which we could not find former applications. Experimentally, the smallest non-zero regularization magnitude we tested was 0.01, which was also the optimal regularization for some conditions and granularities. Even though tangent-FCs together with correlation distance seemed robust to large deviations from optimal magnitudes of regularization, a more exhaustive exploration of regularization on ID rates is needed, in order to make a more comprehensive and generalizable conclusion. In addition, when accounting for fMRI scan length, different intervals of resting state for the same duration should be evaluated, not just the first number of scan volumes.

**ACKNOWLEDGEMENTS**

Data were provided [in part] by the Human Connectome Project, WU-Minn Consortium (Principle Investigators: David Van Essen and Kamil Ugurbil; 1U54MH091657) funded by the 16 NIH Institutes and Centers that support the NIH Blueprint for Neuroscience Research; and by the McDonnell Center for Systems Neuroscience at Washington University.

JG, MD and DK acknowledge financial support from the Indiana CTSI, NIH R21 AA029614 and Indiana Alcohol Research Center P60AA07611, EA acknowledges financial support from the SNSF Ambizione project "Fingerprinting the brain: network science to extract features of cognition, behavior and dysfunction" (grant number PZ00P2_185716).

We thank Nicholas Metcalf and Dr. Sarah Cooley for helping us with the initial processing, the demographics, and the data handling of the validation dataset.

**References:**

1. Abbas, K. *et al.* Geodesic Distance on Optimally Regularized Functional Connectomes Uncovers Individual Fingerprints. *Brain Connectivity* **11**, 333–348 (2021).

2. Fornito, A., Zalesky, A. & Bullmore, E. *Fundamentals of Brain Network Analysis*. (Academic Press, 2016).

3. Sporns, O. *Networks of the Brain*. (MIT Press, 2016).

4. Mantwill, M., Gell, M., Krohn, S. & Finke, C. Brain connectivity fingerprinting and behavioural prediction rest on distinct functional systems of the human connectome. *Commun Biol* **5**, 1–10 (2022).

5. Sripada, C. *et al.* Basic Units of Inter-Individual Variation in Resting State Connectomes. *Sci Rep* **9**, 1900 (2019).

6. Finn, E. S. *et al.* Functional connectome fingerprinting: identifying individuals using patterns of brain connectivity. *Nat Neurosci* **18**, 1664–1671 (2015).

7. Amico, E. & Goñi, J. The quest for identifiability in human functional connectomes. *Sci Rep* **8**, 8254 (2018).

8. Venkatesh, M., Jaja, J. & Pessoa, L. Comparing functional connectivity matrices: A geometry-aware approach applied to participant identification. *NeuroImage* **207**, 116398 (2020).

9. Abbas, K. *et al.* GEFF: Graph embedding for functional fingerprinting. *NeuroImage* **221**, 117181 (2020).

10. Amico, E. *et al.* The disengaging brain: Dynamic transitions from cognitive engagement and alcoholism risk. *NeuroImage* **209**, 116515 (2020).

11. Barnes, A., Bullmore, E. T. & Suckling, J. Endogenous Human Brain Dynamics Recover Slowly Following Cognitive Effort. *PLOS ONE* **4**, e6626 (2009).

12. Fornito, A., Zalesky, A. & Breakspear, M. The connectomics of brain disorders. *Nat Rev Neurosci* **16**, 159–172 (2015).

13. Svaldi, D. O. *et al.* Optimizing differential identifiability improves connectome predictive modeling of cognitive deficits from functional connectivity in Alzheimer's disease. *Human Brain Mapping* **42**, 3500–3516 (2021).

14. van den Heuvel, M. P. & Sporns, O. A cross-disorder connectome landscape of brain dysconnectivity. *Nat Rev Neurosci* **20**, 435–446 (2019).

15. Ponsoda, V. *et al.* Structural brain connectivity and cognitive ability differences: A multivariate distance matrix regression analysis. *Hum Brain Mapp* **38**, 803–816 (2017).

16. Dadi, K., Abraham, A., Rahim, M., Thirion, B. & Varoquaux, G. Comparing functional connectivity based predictive models across datasets. in *2016 International Workshop on Pattern Recognition in Neuroimaging (PRNI)* 1–4 (2016). doi:10.1109/PRNI.2016.7552359.

17. Varoquaux, G., Baronnet, F., Kleinschmidt, A., Fillard, P. & Thirion, B. Detection of Brain Functional-Connectivity Difference in Post-stroke Patients Using Group-Level Covariance Modeling. in *Medical Image Computing and Computer-Assisted Intervention – MICCAI 2010* (eds. Jiang, T., Navab, N., Pluim, J. P. W. & Viergever, M. A.) 200–208 (Springer, 2010). doi:10.1007/978-3-642-15705-9_25.

18. Ng, B. *et al.* Transport on Riemannian Manifold for Functional Connectivity-Based Classification. in *Medical Image Computing and Computer-Assisted Intervention – MICCAI 2014* (eds. Golland, P., Hata, N., Barillot, C., Hornegger, J. & Howe, R.) vol. 8674 405–412 (Springer International Publishing, 2014).

19. Dadi, K. *et al.* Benchmarking functional connectome-based predictive models for resting-state fMRI. *NeuroImage* **192**, 115–134 (2019).

20. Ng, B., Varoquaux, G., Poline, J. B., Greicius, M. & Thirion, B. Transport on Riemannian Manifold for Connectivity-Based Brain Decoding. *IEEE Transactions on Medical Imaging* **35**, 208–216 (2016).

21. Ng, B. *et al.* Distinct alterations in Parkinson's medication-state and disease-state connectivity. *NeuroImage: Clinical* **16**, 575–585 (2017).

22. Rahim, M., Thirion, B. & Varoquaux, G. Population shrinkage of covariance (PoSCE) for better individual brain functional-connectivity estimation. *Medical Image Analysis* **54**, 138–148 (2019).

23. Pervaiz, U., Vidaurre, D., Woolrich, M. W. & Smith, S. M. Optimising network modelling methods for fMRI. *NeuroImage* **211**, 116604 (2020).

24. Dodero, L., Minh, H. Q., Biagio, M. S., Murino, V. & Sona, D. Kernel-based classification for brain connectivity graphs on the Riemannian manifold of positive definite matrices. in *2015 IEEE 12th International Symposium on Biomedical Imaging (ISBI)* 42–45 (2015). doi:10.1109/ISBI.2015.7163812.

25. Wong, E., Anderson, J. S., Zielinski, B. A. & Fletcher, P. T. Riemannian Regression and Classification Models of Brain Networks Applied to Autism. in *Connectomics in NeuroImaging* (eds. Wu, G., Rekik, I., Schirmer, M. D., Chung, A. W. & Munsell, B.) 78–87 (Springer International Publishing, 2018). doi:10.1007/978-3-030-00755-3_9.

26. Qiu, A., Lee, A., Tan, M. & Chung, M. K. Manifold learning on brain functional networks in aging. *Medical Image Analysis* **20**, 52–60 (2015).

27. Simeon, G., Piella, G., Camara, O. & Pareto, D. Riemannian Geometry of Functional Connectivity Matrices for Multi-Site Attention-Deficit/Hyperactivity Disorder Data Harmonization. *Frontiers in Neuroinformatics* **16**, (2022).

28. Congedo, M., Barachant, A. & Bhatia, R. Riemannian geometry for EEG-based brain-computer interfaces; a primer and a review. *Brain-Computer Interfaces* **4**, 155–174 (2017).

29. Yger, F., Berar, M. & Lotte, F. Riemannian Approaches in Brain-Computer Interfaces: A Review. *IEEE Transactions on Neural Systems and Rehabilitation Engineering* **25**, 1753–1762 (2017).

30. Satterthwaite, T. D., Xia, C. H. & Bassett, D. S. Personalized Neuroscience: Common and Individual-Specific Features in Functional Brain Networks. *Neuron* vol. 98 243–245 Preprint at https://doi.org/10.1016/j.neuron.2018.04.007 (2018).

31. Gratton, C. *et al.* Functional Brain Networks Are Dominated by Stable Group and Individual Factors, Not Cognitive or Daily Variation. *Neuron* **98**, 439-452.e5 (2018).

32. Seitzman, B. A. *et al.* Trait-like variants in human functional brain networks. *Proceedings of the National Academy of Sciences of the United States of America* **116**, 22851–22861 (2019).

33. Miranda-Dominguez, O. *et al.* Connectotyping: Model Based Fingerprinting of the Functional Connectome. *PLOS ONE* **9**, e111048 (2014).

34. Van Essen, D. C. *et al.* The WU-Minn Human Connectome Project: an overview. *Neuroimage* **80**, 62–79 (2013).

35. Tolosi, L. & Lengauer, T. Classification with correlated features: unreliability of feature ranking and solutions. *Bioinformatics* **27**, 1986–1994 (2011).

36. Tian, Y. & Zalesky, A. Machine learning prediction of cognition from functional connectivity: Are feature weights reliable? *NeuroImage* **245**, 118648 (2021).

37. Bhatia, R. *Positive definite matrices*. (Princeton university press, 2009).

38. Pennec, X., Fillard, P. & Ayache, N. A Riemannian Framework for Tensor Computing. *Int J Comput Vision* **66**, 41–66 (2006).

39. Blokland, G. A., Mosing, M. A., Verweij, K. H. & Medland, S. E. Twin studies and behavior genetics. *The Oxford handbook of quantitative methods in psychology* **2**, 198–218 (2013).

40. Schaefer, A. *et al.* Local-Global Parcellation of the Human Cerebral Cortex from Intrinsic Functional Connectivity MRI. *Cerebral Cortex* **28**, 3095–3114 (2018).

41. Glasser, M. F. *et al.* A multi-modal parcellation of human cerebral cortex. *Nature* **536**, 171–178 (2016).

42. Marcus, D. S. *et al.* Informatics and data mining tools and strategies for the human connectome project. *Front Neuroinform* **5**, 4 (2011).

43. Glasser, M. F. *et al.* The minimal preprocessing pipelines for the Human Connectome Project. *Neuroimage* **80**, 105–124 (2013).

44. Smith, S. M. *et al.* Resting-state fMRI in the Human Connectome Project. *Neuroimage* **80**, 144–168 (2013).

45. Power, J. D. *et al.* Methods to detect, characterize, and remove motion artifact in resting state fMRI. *Neuroimage* **84**, 320–341 (2014).

46. Amico, E., Arenas, A. & Goñi, J. Centralized and distributed cognitive task processing in the human connectome. *Netw Neurosci* **3**, 455–474 (2019).

47. Cole, M. W., Bassett, D. S., Power, J. D., Braver, T. S. & Petersen, S. E. Intrinsic and task-evoked network architectures of the human brain. *Neuron* **83**, 238–251 (2014).

48. Barachant, A., Bonnet, S., Congedo, M. & Jutten, C. Classification of covariance matrices using a Riemannian-based kernel for BCI applications. *Neurocomputing* **112**, 172–178 (2013).

49. Allen, E. A. *et al.* Tracking whole-brain connectivity dynamics in the resting state. *Cereb Cortex* **24**, 663–676 (2014).

50. Noble, S. *et al.* Influences on the Test–Retest Reliability of Functional Connectivity MRI and its Relationship with Behavioral Utility. *Cerebral Cortex* **27**, 5415–5429 (2017).

51. Cox, R. W. AFNI: software for analysis and visualization of functional magnetic resonance neuroimages. *Comput Biomed Res* **29**, 162–173 (1996).

52. Cox, R. W. & Hyde, J. S. Software tools for analysis and visualization of fMRI data. *NMR in Biomedicine* **10**, 171–178 (1997).

53. Jenkinson, M., Beckmann, C. F., Behrens, T. E. J., Woolrich, M. W. & Smith, S. M. FSL. *NeuroImage* **62**, 782–790 (2012).

54. Tian, Y., Margulies, D. S., Breakspear, M. & Zalesky, A. Topographic organization of the human subcortex unveiled with functional connectivity gradients. *Nat Neurosci* **23**, 1421–1432 (2020).

55. Avants, B. B. *et al.* A reproducible evaluation of ANTs similarity metric performance in brain image registration. *Neuroimage* **54**, 2033–2044 (2011).

56. Patenaude, B., Smith, S. M., Kennedy, D. N. & Jenkinson, M. A Bayesian model of shape and appearance for subcortical brain segmentation. *NeuroImage* **56**, 907–922 (2011).

57. Aquino, K. M., Fulcher, B. D., Parkes, L., Sabaroedin, K. & Fornito, A. Identifying and removing widespread signal deflections from fMRI data: Rethinking the global signal regression problem. *NeuroImage* **212**, 116614 (2020).

58. Koch, I. *Analysis of Multivariate and High-Dimensional Data*. (Cambridge University Press, 2013).

59. Varoquaux, G., Gramfort, A., Poline, J. & Thirion, B. Brain covariance selection: better individual functional connectivity models using population prior. in *Advances in Neural Information Processing Systems* vol. 23 (Curran Associates, Inc., 2010).

60. Barachant, A., Bonnet, S., Congedo, M. & Jutten, C. Multiclass Brain–Computer Interface Classification by Riemannian Geometry. *IEEE Transactions on Biomedical Engineering* **59**, 920–928 (2012).

61. Finn, E. S. & Rosenberg, M. D. Beyond fingerprinting: Choosing predictive connectomes over reliable connectomes. *NeuroImage* **239**, 118254 (2021).

62. Milham, M. P., Vogelstein, J. & Xu, T. Removing the Reliability Bottleneck in Functional Magnetic Resonance Imaging Research to Achieve Clinical Utility. *JAMA Psychiatry* **78**, 587–588 (2021).